# Role of anthropogenic direct heat emissions in global warming

Fei Wang, [a, b] Xingmin Mu, [a, b] Guangju Zhao, [a, b] Peng Gao, [a, b] and Pengfei Li [a]

[a] *Institute of Soil and Water Conservation, Northwest A&F University, Yangling, 712100, Shaanxi, China. E-mail: (Fei Wang)* [wafe@ms.iswc.ac.cn](wafe@ms.iswc.ac.cn); *Fax: +86 29 87012210; Tel: +86 29 87019829*

[b] *Institute of Soil and Water Conservation, Chinese Academy of Sciences and Ministry of Water Resources, Yangling, 712100, Shaanxi, China*

The anthropogenic emissions of greenhouse gases (GHG) are widely realized as the predominant drivers of global warming, but the huge and increasing anthropogenic direct heat emissions (AHE) has not gained enough attention in terms of its role in the warming of the climate system. Based on two reasonable assumptions of (1) AHE eventually transfers to the Earth energy system and (2) the net warming is only driven by the net radioactive forcing (RF) from either GHG or other causes, we analyzed the role of AHE in global warming. The mean annual total AHE of the four main sources including energy consumption, residual heat of electricity generation, biomass decomposition by land use and cover change (LUCC) and food consumption was estimated to be $4.41 \times 10^{20}$ J in 1970-2010, accounting for 6.23% of the net annual heat increase of the Earth reported by IPCC AR5 for the period. The mean annual radioactive forcing (RF) by AHE was up to 29.94 mW/m$^2$ globally in 1981-2010, less than the annual net increase of total GHG (34.80 mW/m$^2$) but more that of $CO_2$ (24.43 mW/m$^2$). The results suggested that AHE played a great role in global warming, urging us to pay more attention to DHE for a better evaluation of RF and more reasonable energy policy including non-GHG energy in the future.

## Background

Warming of the climate system is unequivocal and the anthropogenic emissions of greenhouse gases (GHG) are currently considered as the predominant drivers.[1] The role of the huge and increasing amount of anthropogenic heat emissions (AHE), different from the yearly-fluctuating solar-Earth-energy system in terms of energy source and/or timescale,[2, 3] plays in global warming was rarely evaluated by previous studies on climate system such as IPCC AR5.[1, 2, 3, 4] However, AHE, along with processes of $CO_2$ emission in most cases, can actually influence the heat balance of the Earth, possibly affecting the trajectory of global climate change significantly. There was therefore a need to explore the importance of AHE in the climate warming, especially compared to that of GHG.

# Methodology

Energy consumption, residual heat of electricity generation, biomass decomposition and food consumption are the major sources of AHE. We thus estimated AHE induced by these four anthropogenic activities and discussed their contribution to climate warming in 1971-2010 at a global scale based on two reasonable assumptions of (1) AHE eventually transfers to the Earth energy system and (2) the net warming is only driven by the net radioactive forcing (RF) from either GHG or other sources.

As shown in Table 1, (1) AHE caused by energy consumption was estimated with assuming 10.0% of world's primary energy consumption[5] was released in the forms of motion, light and sound etc. rather than AHE (A reported ratio of un-heating energy consumption was 6.7%)[2]. (2) Residual heat of electricity generation (nuclear, hydroelectricity and renewables), which was not included in energy consumption above was derived based on a thermal equivalence with assuming a 62% of heat lost in modern thermal power stations.[5] (3) AHE from biomass decomposition with land use and cover change (LUCC) was combined with biomass composition based on the $CO_2$ released from LUCC.[6] (4) AHE from food consumption was estimated based on the annual world population in 1971-2010 reported by World Bank, and an average 2,560 kcal/capita/day in 1964-1999 according to FAO report.[7]

Table 1. Mean annual AHE and RFs of AHE and GHG including $CO_2$ in two study periods [a]

| | Caused by | 1971-2010 | 1981-2010 |
|---|---|---|---|
| Mean annual AHE [b] (in $10^{20}$ J) | Energy consumption [b] | 3.08 | 3.35 |
| | Residue of electricity generation | 0.64 | 0.75 |
| | LUCC [c] | 0.49 | 0.49 |
| | Food consumption [d] | 0.21 | 0.22 |
| | **Total** | 4.41 | 4.82 |
| Mean annual radioactive forcing(RF) (in mW/m$^2$) | Total AHE | 27.42 | 29.94 |
| | Total GHG (relative to 1750) | - | 2326.00 |
| | Total GHG (Net increase) [e] | - | 34.80 |
| | $CO_2$ (Net increase)[e] | - | 24.43 |

Notes:

[a]: We used two periods for comparing AHE with global heat increase reported by IPCC AR5 in 1971-2010,[1] and comparing RF caused by AHE, total GHG and $CO_2$ in 1981-2010.[8]

[b]: The accelerating geothermal emission during the production of the fossil fuels and other minings excluded.

[c]: The heat released from biomass decomposition assuming a rate of 37000 j/(g carbon), while the less solar energy fixed by plant and evapotranspiration induced by LUCC were excluded.

[d]: The shares of animal and plant foodstuff and the heat for animal foodstuff production converted via feed conversion ratio were not considered.

[e]: Like IPCC's relative to 1750, annual net increase of RF from total GHG and $CO_2$ was abstracted as a year-by-year relative.[5]

## Results and discussions

The mean annual total AHE was $4.41\times10^{20}$ J in 1970-2010, and it was mainly contributed by energy consumption (Table 1). As reported by IPCC AR5, "more than 60% of the net energy increase in the climate system is stored in the upper ocean (0–700 m)" and "the increase in upper ocean heat content during this time period estimated from a linear trend is likely $17\times10^{22}$ J",[1] the whole net heat increase of the Earth was no more than $28.3\times10^{22}$ J (e.g. $26.1\times10^{22}$ J when 65% employed), equivalent to an average annual heat increase of $70.75\times10^{20}$ J, in 1971-2010. Thus, AHE accounted for 6.23% of the net annual heat increase of the Earth in 1971-2010. Because we did not consider the AHE from many other sources (see footnotes of Table 1) and a high value of the net heat increase in the Earth ($28.3\times10^{22}$ J) was used in our calculation, the contribution of AHE estimated here was likely to be an underestimate.

The mean annual RF caused by AHE was 29.94 mW/m[2] globally in 1981-2010. It was close to mean annual net increase of RF caused by total GHG (34.80 mW/m[2]) and $CO_2$ (24.43 mW/m[2]). According to Assumption 2, the contribution of AHE to global warming was 13.97% less and 22.55% more than that of total GHG and $CO_2$ respectively. Hence, contribution of AHE is much greater than the estimation of "only about 1% of the energy flux being added to Earth because of anthropogenic greenhouse gases".[3]

Total anthropogenic RF (including GHG, short lived gases and aerosols, etc.) in 2011 (2.29 W/m[2]) relative to 1750[1] was 18.94% less than RF from GHG in 2011 (2.825 W/m[2]) relative to 1750.[8] It was inferred that total anthropogenic RF was less than RF by AHE given the latter was closer to RF by GHG, **an unavoidable paradox** emerges: (1) If the net global warming was a result of the total anthropogenic RF[1], did the AHE vanish abruptly? Or (2) if assuming the total anthropogenic RF keeping stable since 2000, would AHE caused RF drive the Earth to be warmer because the total anthropogenic RF was less than RF from AHE? If so, what will be the total human impact?

## Conclusions

Based on our assumptions and estimations, the impact of AHE on global warming should receive more attention because of its great contribution. With social and technology development of our globe, we will undoubtedly need more energy. It is therefore necessary to combine this forcing term (AHE) with widely agreed forcing of GHG and other factors for a better understanding of climate change. It is also necessary to re-think about current environment and energy (including non-GHG energy) policies and technologies in order to mitigate global warming in the future.

## Acknowledgements

The work was partially supported by NSFC (grant no. 41171420), the Key Research Program of the Chinese Academy of Sciences (grant no. KZZD-EW-04), the National Key Basic Research Special Foundation of China (grant no. 2014FY210100) and the West Light Foundation of the Chinese Academy of Sciences.